\documentclass[]{aastex631}

\usepackage[version=3]{mhchem}

\begin{document}

\title{Distinguishing oceans of water from magma on mini-Neptune K2-18b}

\author[0000-0002-8713-1446]{Oliver Shorttle}
\affiliation{Institute of Astronomy, University of Cambridge, UK}
\affiliation{Department of Earth Sciences, University of Cambridge, UK}

\author[0000-0002-2828-0396]{Sean Jordan}
\affiliation{Institute of Astronomy, University of Cambridge, UK}

\author[0000-0002-8368-4641]{Harrison Nicholls}
\affiliation{Atmospheric, Oceanic and Planetary Physics, University of Oxford, UK}

\author[0000-0002-3286-7683]{Tim Lichtenberg}
\affiliation{Kapteyn Astronomical Institute, University of Groningen, The Netherlands}

\author[0000-0002-0673-4860]{Dan J. Bower}
\affiliation{Center for Space and Habitability, University of Bern, Switzerland}


\begin{abstract}

Mildly irradiated mini-Neptunes have densities potentially consistent with them hosting substantial liquid water oceans (`Hycean' planets).  The presence of \ce{CO2} and simultaneous absence of ammonia (\ce{NH3}) in their atmospheres has been proposed as a fingerprint of such worlds.  JWST observations of K2-18b, the archetypal Hycean, have found the presence of \ce{CO2} and the depletion of \ce{NH3} to $<100\,\text{ppm}$; hence, it has been inferred that this planet may host liquid water oceans.  In contrast, climate modelling suggests that many of these mini-Neptunes, including K2-18b, may likely be too hot to host liquid water.  We propose a solution to this discrepancy between observation and climate modelling by investigating the effect of a magma ocean on the atmospheric chemistry of mini-Neptunes.  We demonstrate that atmospheric \ce{NH3} depletion is a natural consequence of the high solubility of nitrogen species in magma at reducing conditions; precisely the conditions prevailing where a thick hydrogen envelope is in communication with a molten planetary surface.  The magma ocean model reproduces the present JWST spectrum of K2-18b to $\lesssim 3\,\sigma$, suggesting this is as credible an explanation for current observations as the planet hosting a liquid water ocean.  Spectral areas that could be used to rule out the magma ocean model include the $>4\,\mu\text{m}$ region, where \ce{CO2} and \ce{CO} features dominate:  Magma ocean models suggest a systematically lower \ce{CO2}/\ce{CO} ratio than estimated from free chemistry retrieval, indicating that deeper observations of this spectral region may be able to distinguish between oceans of liquid water and magma on mini-Neptunes.

\end{abstract}

\section{Introduction} \label{sec:intro}

K2-18b, a mini-Neptune class exoplanet for which no analogue exists in the Solar System, has an observed mass, radius, and atmospheric transmission spectrum that indicates it possesses an \ce{H2}-dominated atmosphere but an unknown interior composition \citep{Benneke2019,2019NatAs...3.1086T}. With a mass of 8.6 M$_{\oplus}$ and a radius of 2.6 R$_{\oplus}$ \citep{Benneke2019}, K2-18b has a bulk density in between that of the Earth and Neptune that is consistent with a range of possible interior structures. Previous studies have used interior structure models to find what mass fractions of H/He envelope are required to explain the observed radius if the mass is (a) dominated by an Fe-core and silicate mantle, (b) resembles a Neptune-like interior consisting of an Fe-rich nucleus and ice-rich gaseous envelope, or (c) is instead dominated by \ce{H2O} phases \citep{Madhusudhan2020, Nixon2021}. In the latter, \ce{H2O}-dominated case, a minimal core mass fraction of 10\% of the total mass budget of the planet would enable a water mass fraction of $\sim$\,90\% and an envelope mass fraction of only 0.006\% to fit the observed radius. This scenario is proposed to be admissible to the existence of a liquid water ocean in contact with the thin \ce{H2}-dominated atmosphere \citep{Madhusudhan2020}. If it were possible for the liquid water phase to remain stable under these conditions, then the small radius of K2-18b's M-dwarf host star, and the large scale height of its \ce{H2}-dominated atmosphere, would enable canonically habitable conditions on K2-18b to be identifiable by transmission spectroscopy with JWST.  Such a scenario would mark an important extension and test of the traditional `liquid water' habitable zone concept to the geologically uncharted territory of so called `Hycean' (Hydrogen-atmosphere with water-ocean) exoplanets in general \citep{Madhu-Hyceans-2021,2021Univ....7..172S,2014ApJ...784...63H,2016AsBio..16...89C}.  
\\\\
Characterisation of these mini-Neptunes has now begun.  In the recent transmission spectrum of K2-18b obtained by JWST, the presence of \ce{CO2} and simultaneous non-detection of \ce{NH3} has been linked to the possibility of a liquid water ocean underlying the \ce{H2}-dominated atmosphere \citep{Madhusudhan2023}. While the atmospheric scale height makes spectral features amenable to detection with transmission spectroscopy, the pressure level probed is $\lesssim$\,0.1 bar and hence the surface remains out of reach to remote observations. Uncertainty therefore remains as to whether an (a) rock-dominated, (b) Neptune-like, or (c) \ce{H2O}-dominated composition best represents the true interior structure of K2-18b. 
\\\\
Methods of separating out the degenerate interior structure models have been offered based on the atmospheric chemistry of mini-Neptunes.  Diagnostic chemical indicators have been proposed for when liquid-water solubility equilibria are included in the thermochemical equilibria, compared with when only pure gas-phase equilibria are considered \citep{Hu2021}: \ce{CO2} and \ce{NH3} being key species in this regard. In gas-phase thermochemical equilibrium, \ce{NH3} is the dominant carrier of N atoms at high pressure and temperature conditions in a deep \ce{H2}-rich envelope.  The expectation is that this \ce{NH3} would be dynamically transported to the observable regions of the atmosphere if K2-18b possessed a Neptune-like interior \citep{Yu2021, Tsai2021}. Alternatively, the solubility of \ce{NH3} in liquid water would enable it to be efficiently sequestered if a liquid water ocean were present, and thus it would be depleted in the atmosphere to levels below detectability by transmission spectroscopy \citep{Hu2021}.  Liquid water may also manifest through the \ce{CO2} abundance of thin atmospheres overlying water.  In this case, the significant storage capacity of liquid water for \ce{CO2} means that in low-mass atmospheres \ce{CO2} is buffered by the oceans and \ce{CO2} is potentially able to dominate over \ce{CO} and \ce{CH4} \citep{Hu2021}.
\\\\
A free-chemistry retrieval of the atmospheric transmission spectrum JWST obtained of K2-18b --- i.e., a fit of the atmospheric chemistry to the spectrum without constraint by thermochemical equilibrium or photochemical kinetics --- reported a model preference for $<100$\,ppm \ce{NH3} at a 95\% confidence level \citep{Madhusudhan2023}.  The ammonia depletion, combined with the detection of \ce{CO2}, were taken by \citet{Madhusudhan2023} to suggest that liquid-water solubility equilibria may indeed be depleting \ce{NH3} from K2-18b's observable atmosphere.
\\\\
Radiative-convective model developments, however, suggest that if K2-18b possesses a significant inventory of \ce{H2O} and an \ce{H2}-dominated atmosphere, then it would lie inside of the inner edge of the liquid water habitable zone.  These climate models predict atmospheres for K2-18b consisting of a supercritical \ce{H2}-\ce{H2O} mixture in runaway greenhouse \citep{Innes2023}, rather than the proposed state of a stable liquid water ocean beneath a \ce{H2}-dominated atmosphere. K2-18b receives a similar instellation flux from it's host star as the Earth receives from the Sun, however methods to model the warming effects of \ce{H2}-dominated atmospheres for mildly irradiated mini-Neptunes are currently still being developed \citep{Koll2019, 2021JGRE..12606711L, Innes2023, Pierrehumbert2023}. In particular, the low mean molecular weight of \ce{H2} as a background gas in an atmosphere containing a significant condensible component leads to qualitatively different thermodynamic heating behaviours than those considered in estimations of the canonical liquid water habitable zone \citep{Guillot95,Leconte17}; where heavier background gases such as \ce{N2} dominate \citep{Koll2019}. In the scenario proposed for K2-18b as a Hycean world, the condensible component (\ce{H2O}) is heavier than the background gas (\ce{H2}) thus the decrease in temperature with altitude leads to a sharp decrease in atmospheric mean molecular weight. By accounting for this compositional gradient recent radiative-convective modelling has shown that convective inhibition results in temperatures far higher than otherwise predicted, rendering the majority of the known population of mini-Neptune class exoplanets, including K2-18b, inside of this revised inner edge of the habitable zone \citep{Innes2023}. This model development is difficult to reconcile with the observed depletion of \ce{NH3} in the transmission spectrum of K2-18b if the depletion is due to solubility in a liquid water ocean in contact with the \ce{H2}-dominated atmosphere.
\\\\
One straightforward interpretation of the low atmospheric \ce{NH3} abundance would be that it indicates a planet with inherently low bulk N.  Whilst this is hard to rule out, especially given a lack of population-level data on mildly irradiated mini-Neptunes, we think it is unlikely to be an adequate explanation for the K2-18b observations for two reasons.  First, this does not explain why there should be appreciable \ce{CO2} in the atmosphere, which has itself been suggested as consistent with shallow atmospheres above liquid water oceans \citep{Hu2021}.  Second, there is substantial evidence from the solar system \citep{
,grewal2021rates,grewal2023origin,chen2022impact} and astrochemical studies \citep{bergin2015tracing,2021PhR...893....1O} that a significant fraction of nitrogen is carried in refractory phases, and will be delivered to growing planets alongside carbon and other volatiles \citep[][]{2023ASPC..534.1031K,suer2023_frontearthsci}.  It is therefore unclear why such a large depletion in nitrogen should be found in a planet that from mass-radius constraints alone is required to have a significant volatile-element fraction.
\\\\
Instead, we pursue a resolution to the apparent discrepancy between observation and climate modelling for K2-18b: that the presence of \ce{CO2}, and non-detection of \ce{NH3}, in its atmosphere is consistent with an \ce{H2}-dominated atmosphere underlain by a liquid ocean, not of water, but of silicate magma. The abundances of volatile species in an atmosphere in contact with silicate magma is set by the solubility equilibria of molecules dissolving in the silicate melt. The solubility of nitrogen in silicate systems has been extensively studied to investigate the origin of nitrogen in the terrestrial and Venusian atmospheres, and in the mantles of differentiated solar system bodies \citep[][and references therein]{dasgupta2022,suer2023_frontearthsci}. Laboratory experiments have enabled parameterisations of nitrogen solubility to be formulated over a wide range of pressures, temperatures, and oxygen fugacities. These studies have demonstrated that, under oxidising conditions, nitrogen physically dissolves as \ce{N2} molecules into cavities of the silicate network according to Henry's law \citep{Libourel2003}. However, at more reducing conditions, below the iron-wustite buffer, nitrogen chemically dissolves as \ce{N^{3-}} ions by forming complexes with atoms in the silicate melt network, increasing its solubility by up to 5 orders of magnitude \citep{Libourel2003, dasgupta2022}. In the context of exoplanets such as K2-18b, while \ce{NH3} remains a dominant carrier of N atoms at the base of the atmosphere in thermochemical equilibrium, nitrogen solubility in silicate melt can nonetheless result in a very low atmospheric nitrogen content \citep{wordsworth2022atmospheres}. Nitrogen solubility in a magma ocean could therefore mimic the effect of a shallow surface inhibiting the recycling of \ce{NH3} from thermochemical equilibrium in a deep gaseous envelope. 
\\\\
Motivated by this possibility, in this paper we investigate whether the magma ocean scenario can lead to the observed presence of \ce{CO2} and depletion of \ce{NH3} in the atmosphere of K2-18b. Section \ref{sec:methods} outlines the modelling tools that we use and the parameter space that we explore for the magma ocean scenario. Section \ref{sec:results} presents our results for the atmosphere - magma ocean system, and compares transmission spectra of the model atmospheres to the observed JWST transmission spectrum of K2-18b. In section \ref{sec:discussion} we discuss the magma ocean and water ocean interpretations of K2-18b, and we conclude in section \ref{sec:conclusions}.

\section{Modelling K2-18\MakeLowercase{b} as a magma ocean \label{sec:methods}}

\begin{figure}[htbp]
    \centering
    \includegraphics[width=0.8\columnwidth]{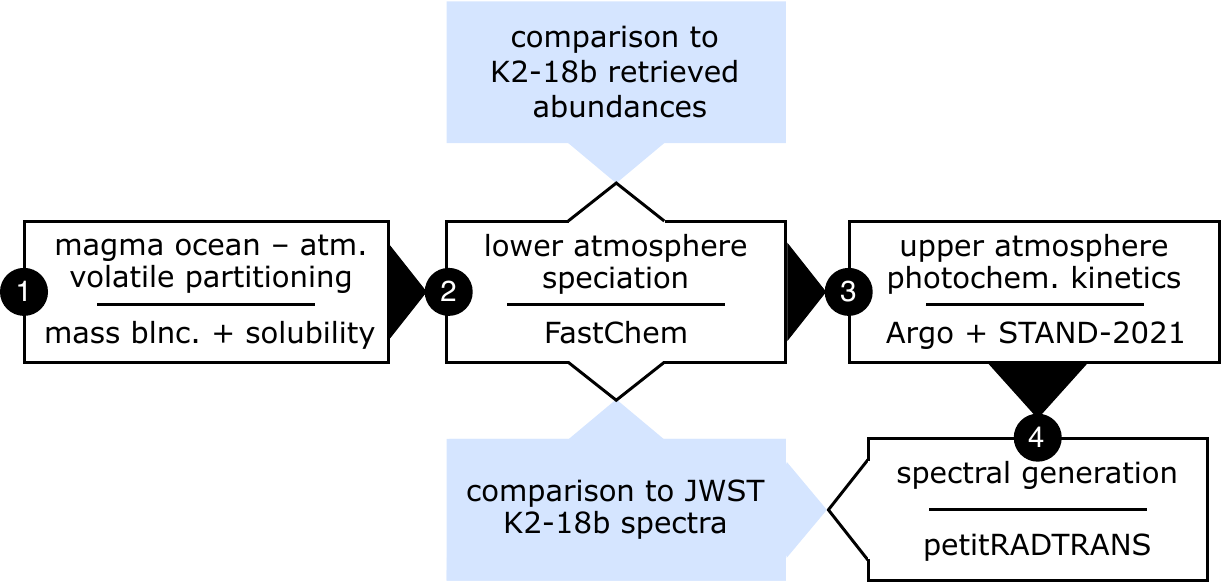}
    \caption{An outline of the methods followed to generate model atmospheric compositions and spectra of K2-18b as a magma ocean.}
    \label{fig:method_schematic}
\end{figure}

An outline of the method we follow is shown in Figure \ref{fig:method_schematic} and details of the steps are given in the sections below. To summarise:  At the heart of the approach is the magma ocean -- atmosphere equilibration (step 1), where volatile mass is distributed between the molten interior and atmosphere/envelope of the planet.  Critically for the ensuing atmospheric chemistry, this distribution is governed by the solubility of volatile elements in the magma ocean. This first step dictates the composition of the atmosphere, but can necessarily only consider solubility of certain key species that have experimental constraints.  In step 2, we calculate pure gas phase thermochemical equilibrium of this predicted atmosphere at its base, directly overlying the magma ocean.  This allows us to fully speciate the atmosphere for an equilibrium chemistry solution to what K2-18b's atmosphere should look like underlain by a magma ocean, and also is essential input for step 3. At this point we can also make the first comparison to the atmosphere of K2-18b, here comparing broadly to the observed transmission spectrum and retrieved molecular abundances.  In step 3 we perform a simple characterisation of the effect of atmospheric structure (pressure temperature profile) and photochemical kinetics on the atmosphere predicted from magma ocean -- atmosphere equilibrium.  This step tests whether the heterogeneous equilibria between gas and magma, set at the atmosphere's base, survives to the $\sim{1\,\text{mbar}}$ level transmission spectroscopy probes.  Step 4 then calculates the transmission spectrum of these atmospheres for direct comparison to the JWST spectra of K2-18b. 
\\
\\
\subsection{Equilibrium outgassing of a magma ocean}
Volatile abundances in the melt and in the overlying atmosphere are set by their solubility and equilibrium chemistry, respectively. This is achieved by requiring mass conservation between the two inventories \citep{2022PSJ.....3...93B}. The concentrations of \ce{H2O}, \ce{CO2}, \ce{CH4}, and \ce{N2} in the melt are set by their partial pressures in the atmosphere according to the solubility laws derived by \citet{sossi2023}, \citet{dixon1995}, \citet{Ardia2013SolubilityOC}, and \citet{dasgupta2022} respectively. The solubility of these volatiles depends on the oxygen fugacity (measured relative to the Iron-W\"ustite buffer), and temperature of the melt \citep{suer2023_frontearthsci}. Given a total mantle mass, this provides us with the total mass of dissolved \ce{H}, \ce{C}, \ce{O}, and \ce{N}. The three reactions \mbox{\ce{CO2 + 2 H2 <=> CH4 + O2}}, \mbox{\ce{2 CO2 = 2 CO + O2}}, and \mbox{\ce{2 H2O = 2 H2 + O2}} are assumed to attain thermochemical equilibrium, providing us with the partial pressures of \ce{CH4}, \ce{CO}, and \ce{H2} respectively. The total masses of \ce{H}, \ce{C}, \ce{O}, and \ce{N} in the atmosphere are obtained from the volatile partial pressures \citep{2022PSJ.....3...93B}.
\\\\
For high planetary volatile inventories the pressures at the base of the atmosphere, at the interface with the magma ocean, can be high (up to several gigapascal).  Under these conditions the atmosphere will not behave as an ideal gas.   In this work we have chosen the solubility laws most appropriate for the conditions being modelled, which find a compromise between fit to pressure/temperature conditions (for \ce{CH4} and N, \citet{Ardia2013SolubilityOC,dasgupta2022}), and compositional conditions (for \ce{CO2} and \ce{H2O} \citet{dixon1995,sossi2023}).  In terms of speciating the atmosphere, the chosen solubility laws allow us to estimate the atmospheric composition due to magma ocean solubility effects, and then calculate its full speciation at low pressure where assumptions of ideality hold.  
\\\\
To fully speciate the gas-phase chemistry of the atmosphere from the minimal species set used for solubility calculations, the atmospheric composition calculated above is passed to FastChem \citep{stock2022fastchem,kitzmann2023_mnras}.  FastChem calculates the gas-phase thermochemical equilibrium in the atmosphere at the prescribed pressure and temperature.  This then provides a full set of minor gas species to go forward for photochemical kinetic modelling of the atmospheric chemical structure above the magma ocean interface.
\\\\
Magma ocean calculations were run over a wide range of parameters given in Table \ref{tab:params}, with solar metallicity ratios using data from \citet{Asplund2009Solar}.  Surface temperature, the effective magma ocean mass (via `mantle mass'), how oxidising the magma ocean is (relevant for nitrogen solubility), the hydrogen budget of the planet and the C/H ratio of the planet were all varied.  The planet's nitrogen budget was also varied, by scaling the solar N/H ratio, over four orders of magnitude.   These choices of compositional parameters lead to total abundances (atmosphere  plus magma ocean) of carbon ranging from to $10^{-8}$ to 3.8\,wt\% and of nitrogen ranging from $10^{-3}$ to $10$\,ppm (expressed as a mass fraction of the mantle).  We do not show the results of models predicting atmospheres with basal pressures exceeding $10^5\,\text{bar}$, as above these pressures little experimental data exist on volatile partitioning into magmas.  For the parameters shown in Table \ref{tab:params}, fewer than 1\% of models predict such high pressure atmospheres. 

\begin{table}
\begin{center}
\caption{Parameters used for magma ocean modelling.\label{tab:params}}
\begin{tabular}{l l l}
\\\hline
Parameter & Units & Values \\\hline\hline
\multicolumn{3}{l}{\emph{Independent parameters}}\\
Mantle Mass & $M_{planet}$ & $\{0.001, 0.01, 0.1, 1\}$\\
Surface temperature & K & $\{1500, 2000, 2500, 3000\}$\\
$f\text{O}_2$ (mantle) & $\Delta\text{IW}$ & $\{-5, -2, 0, 2, 4\}$\\
Hydrogen budget & ppm (by mass of mantle) & $\{1, 10, 100, 1000, 10000\}$\\
C/H & $(\text{C/H})_{solar}$ & $\{0.01, 0.05, 0.1, 1, 10, 100\}$\\
\multicolumn{3}{l}{\emph{Dependent parameters}}\\
Nitrogen budget & ppm (by mass of mantle) & $=\text{(N/H)}_{solar}\times\left(\text{H}\,\text{(ppm)}\right)$\\
\hline
\end{tabular}
\end{center}
\end{table}

\subsection{Atmospheric structure}

A subset of the gas-phase chemistry obtained above, from the atmosphere-magma ocean equilibrium model, is then used as a lower boundary condition to generate chemical profiles for the full atmosphere using the photochemical-kinetics code ARGO \citep{Rimmer2016,Rimmer2019,Rimmer2021}. ARGO solves for the chemical profiles as a function of pressure throughout the atmosphere, from the surface pressure determined by the atmosphere-magma ocean system, up to a pressure of 10$^{-8}$\,bar at the top of the atmosphere. ARGO treats H/C/N/O chemistry self-consistently over 100--30,000\,K \citep{Rimmer2016,Hobbs2021}, and has been applied to both giant exoplanet atmospheres \citep[e.g.,][]{Tsai2023} and rocky exoplanet atmospheres \citep[e.g.,][]{Jordan2021}. Many other elements, such as S/P/Cl and other heavy metals, are also included in the network but do not feature in the work presented here as only H/C/N/O chemistry is treated in the magma ocean model.
\\\\
ARGO accounts for atmospheric thermochemistry, photochemisty, vertical dynamical transport, and condensation, by solving the set of reactions listed in the chemical network STAND-2021, as a system of coupled non-linear differential equations. At every altitude/pressure level, ARGO solves the continuity equation:
\begin{equation}
\dfrac{dn_{\rm X}}{dt} = P_{\rm X} - L_{\rm X}n_{\rm X} - \dfrac{\partial \Phi_{\rm X}}{\partial z},
\label{eq:argo_eqn}
\end{equation}
where $n_{\ce{X}}$ (cm$^{-3}$) is the number density of species $\ce{X}$, $P_{\rm X}$ (cm$^{-3}$ s$^{-1}$) is the rate of production of species \ce{X}, $L_{\rm X}$ (s$^{-1}$) is the rate constant for loss of species \ce{X}, and $\partial \Phi_{\rm X}/\partial z$ (cm$^{-3}$ s$^{-1}$), the divergence of the vertical diffusion flux, accounts for eddy- and molecular-diffusion.
\\\\
The system of reactions is solved at each pressure level for a timescale determined by the eddy diffusion profile, which parameterises vertical transport in a 1D model. We use an eddy diffusion profile equal to 10$^{6}$\,cm$^2$s$^{-1}$ in the lower atmosphere up to a pressure of $\sim$ 1\,bar, then varying with the inverse square root of the atmospheric pressure up to a maximum value of 10$^{11}$\,cm$^2$s$^{-1}$ \citep{Moses2022}. The chemical profiles are relatively insensitive to the assumed value of the eddy diffusion profile in the lower atmosphere as reactions rates are fast in the hot deep atmosphere (see below for discussion of thermal structure) and only become slow and quenched in the cold upper regions of the atmosphere.
\\\\
The rate of thermochemical reactions are determined by the pressure, temperature, and chemical composition at a given pressure level. The temperature profile that we use for this is taken from previous radiative-convective modelling of the atmosphere of K2-18b \citep{Benneke2019}, which provides a temperature structure between $\sim 10^{-8}$ -- 4\,bar. For atmospheric models that go to deeper pressures we extrapolate the temperature profile linearly in log(P) space, down to the desired pressure. This is an approximation of what the true temperature structure of K2-18b would look like and future work using coupled models, that treat both radiative-convective equilibrium and photochemical-kinetics, would be required to further investigate how the precise atmospheric temperature structure can modulate the chemistry. Such coupled climate-chemistry models are computationally very intensive and thus not suitable for ensemble grids covering a wide multidimensional parameter space.  The approximate temperature profile that we adopt is sufficient for our purposes of simply demonstrating that the presence of \ce{CO2} and absence of \ce{NH3} due to magma ocean -- atmosphere equilibration is propagated to the observable regions of the atmosphere; i.e., photochemical kinetics do not reverse or overprint the thermochemistry.  We are not aiming to find the precise atmospheric structure to fit the observed spectrum of K2-18b.
\\\\
The rate of photochemical reactions are determined by the flux profile incident at the top of the atmosphere. For this, we use the stellar spectrum of the M2.5 star GJ176, following \citet{Scheucher2020}. GJ176 is similar in its stellar properties to K2-18 and is readily available from the MUSCLES database \citep{france2016muscles,youngblood2016muscles,loyd2016muscles}. Additionally, water condensation is treated in ARGO by converting \ce{H2O}\,(g) to \ce{H2O}\,(l) when the partial pressure of \ce{H2O}\,(g) is in excess of the saturation vapour pressure. Using the temperature pressure profile prescribed by \citep{Benneke2019}, this occurs at $\sim$ 1\,bar in our models; i.e., below the depth probed by transmission spectroscopy.

\subsection{Spectral modelling}

We compare our results to the transmission spectrum observed for K2-18b from JWST by calculating model transmission spectra from the atmosphere -- magma ocean system using petitRADTRANS. petitRADTRANS is an open source, radiative transfer package built for the spectral characterisation of exoplanet atmospheres (see \citet{Molliere2019} for a full description of the code). The code has been used amongst the community extensively, particularly for the case of \ce{H2}-dominated atmospheres, to generate transmission spectra, emission spectra, and for the retrieval of atmospheric parameters from observational data. We consider all opacity sources due to prominent HCNO-molecules and continuum opacities included in the petitRADTRANS opacity database. The important opacity sources in this work are due to molecular absorption by \ce{CH4}, \ce{H2O}, \ce{NH3}, \ce{CO2}, \ce{CO}, and continuum opacity due to \ce{H2} collision induced absorption. The amplitude of spectral features in transmission spectroscopy depends on the atmospheric scale height, H$_{sc} = kT/\mu g$ (where $k$ is the Boltzmann constant, $T$ is temperature, $\mu$ is mean molecular weight of the atmosphere, and $g$ is gravitational acceleration). The temperature structure of \citet{Benneke2019} was used for self-consistency with the atmospheric model, and a value of 12.43\,ms$^{-2}$ was used for the gravity \citep{Benneke2019}. The mean molecular weight was calculated self-consistently for each model atmosphere based on all species' mixing ratios.
\\\\
We use petitRADTRANS at two stages in our analysis: first, transmission spectra are calculated directly from the output of the atmosphere - magma ocean equilibrium system (step 2 in figure \ref{fig:method_schematic}). At this stage, we assume a homogeneous composition throughout the atmosphere. The mean molecular weight and individual species' mass fractions are calculated from the respective contributions of each species' mixing ratio. We compare the output transmission spectra to the observed spectrum of K2-18b and select a sample of the best fitting models to further test with the more computationally intensive photochemical-kinetics model, described above. Using the output of the atmosphere -- magma ocean system as a lower boundary condition, the photochemical-kinetics model solves for inhomogeneous atmospheric composition allowing for atmospheric chemistry and disequilibrium chemistry throughout the atmosphere. We then use petitRADTRANS to model transmission spectra for this set of full atmosphere models (step 4 in figure \ref{fig:method_schematic}), now calculating the individual species' mass fractions and the atmospheric mean molecular weight at every pressure level in the atmosphere based on the chemical profiles obtained. At this stage, a grey cloud was additionally included over a range of pressure levels, and a reduced chi squared fit to the observation data from \citet{Madhusudhan2023} was calculated to obtain a set of best fitting models under the magma ocean scenario.

\section{Results \label{sec:results}}

\subsection{Atmospheric composition}
\begin{figure}[htbp]
    \centering
    \includegraphics[width=0.7\columnwidth]{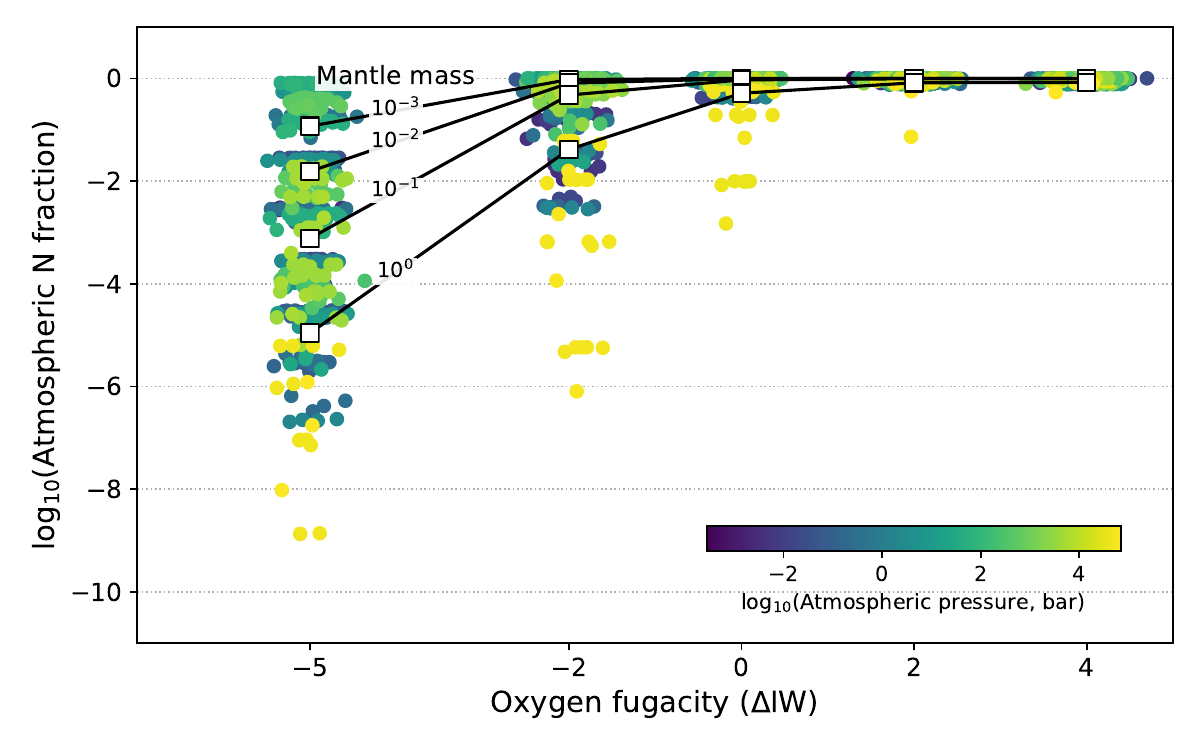}
    \caption{The fraction of nitrogen in the planet's atmosphere compared to the planet's total inventory, as a function of magma ocean oxygen fugacity.  As oxygen fugacity is decreased, nitrogen's increased solubility depletes the atmosphere by orders of magnitude.  Each coloured circle represents a model run for a given set of parameters (Table \ref{tab:params}); symbols have been colored by total atmospheric pressure (bar), and have been randomly displaced in the x direction for ease of visualisation.  White squares are averages for runs of a given mantle mass (expressed as fractional mass of the planet) at a particular oxygen fugacity, and illustrate the important role mantle (or equivalently magma ocean) mass has in enabling significant N depletion.}
    \label{fig:n_budget}
\end{figure}

How oxidising a magma is has a profound effect on the solubility of nitrogen.  The effect of this on the modelled magma ocean atmospheres of K2-18b are shown in Figure \ref{fig:n_budget}, which shows results for all models run. Over the 9 orders of magnitude of oxygen fugacity investigated, the planet's nitrogen budget in the atmosphere in general decreases by at least 1 order of magnitude, and in most cases decreases by a factor of 1000 or more.  Figure \ref{fig:n_budget} makes clear that these most extreme depletions are made possible in an atmosphere overlying a large magma ocean mass at the most reducing conditions.  Whilst not directly commenting on the abundance of spectrally active species like \ce{NH3}, such large drops in atmospheric nitrogen must inevitably lead to large depletions in nitrogen species, even those thermodynamically favoured to be the major atmospheric N reservoirs.  
\\\\
This simple result illustrates the significance of considering volatile solubility when predicting the composition of atmospheres overlying magma oceans.  We investigate the atmospheric chemical and observational consequences in the subsequent sections.

\subsection{Atmospheric chemical abundances}
\begin{figure}[htbp]
    \centering
    \includegraphics[width=\columnwidth]{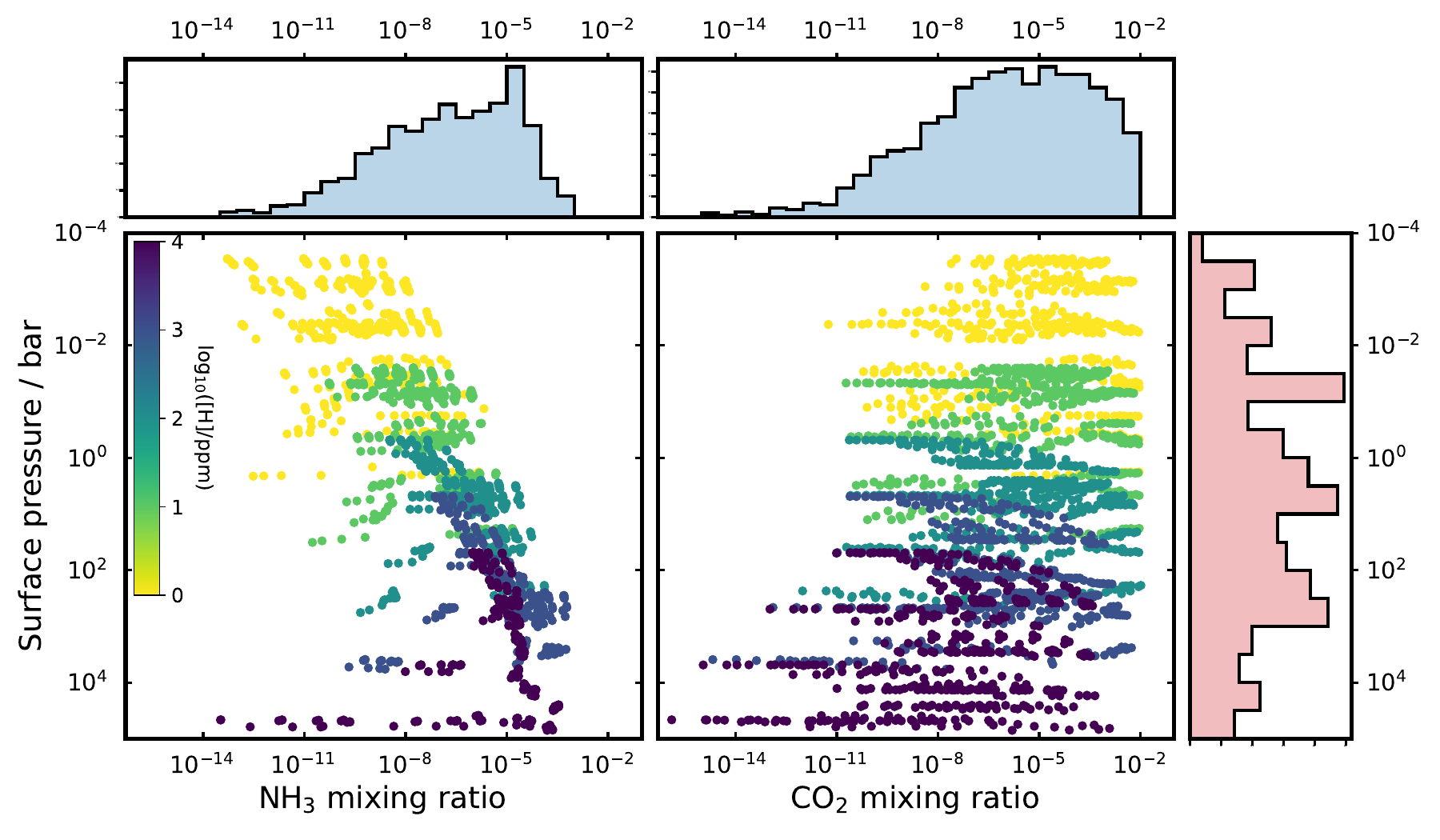}
    \caption{Total atmospheric surface pressure (bar) and volume mixing ratio of \ce{NH3}, and \ce{CO2} output from the atmosphere-magma ocean system over the parameter space which we have sampled, outlined in table \ref{tab:params}. This corresponds to the output of step 2 in the schematic presented in figure \ref{fig:method_schematic}. Points are colour-coded by the log of the planetary hydrogen budget, in units part-per-million of the planetary mass. Histograms of the distribution of \ce{NH3} mixing ratios (top left), \ce{CO2} mixing ratios (top right) and total surface pressures (lower right) from across the model runs are shown alongside.}
    \label{fig:nh3}
\end{figure}

The depletion of \ce{NH3} and the presence of significant \ce{CO2} in an \ce{H2}-dominated atmosphere underlain by a magma ocean can be achieved over a variety of surface pressures, ranging from tenuous atmospheres to thick envelopes. Figure \ref{fig:nh3} demonstrates the range of \ce{NH3} mixing ratios, \ce{CO2} mixing ratios, and surface pressures that result from atmosphere-magma ocean equilibrium over the parameter space that we have sampled (Table \ref{tab:params}). The surface pressures of the resulting atmospheres range from $\sim$\,10$^{-4}$ -- 10$^{8}$\,bar, and the \ce{NH3} mixing ratios range from $\sim$\,10$^{-18}$ -- 10$^{-3}$, across the parameter space. For K2-18b specifically, in the limiting case of a pure silicate interior and no heavy Fe-rich core fraction, an \ce{H2} envelope of $\gtrsim$\,10$^{3}$\,bar would be required to explain the observed radius \citep{Madhusudhan2020}. Our results demonstrate that this is achieved for a wide range of parameter space in the magma ocean scenario, and can thus explain the observed mass, radius, and \ce{NH3} depletion for a range of interior compositions that include an Fe-rich core and silicate mantle of varying core-mantle mass fractions. 
\\
\\
\\
\subsection{Magma ocean fit to transmission spectra}
\begin{figure}[htbp]
    \centering
    \includegraphics[width=\columnwidth]{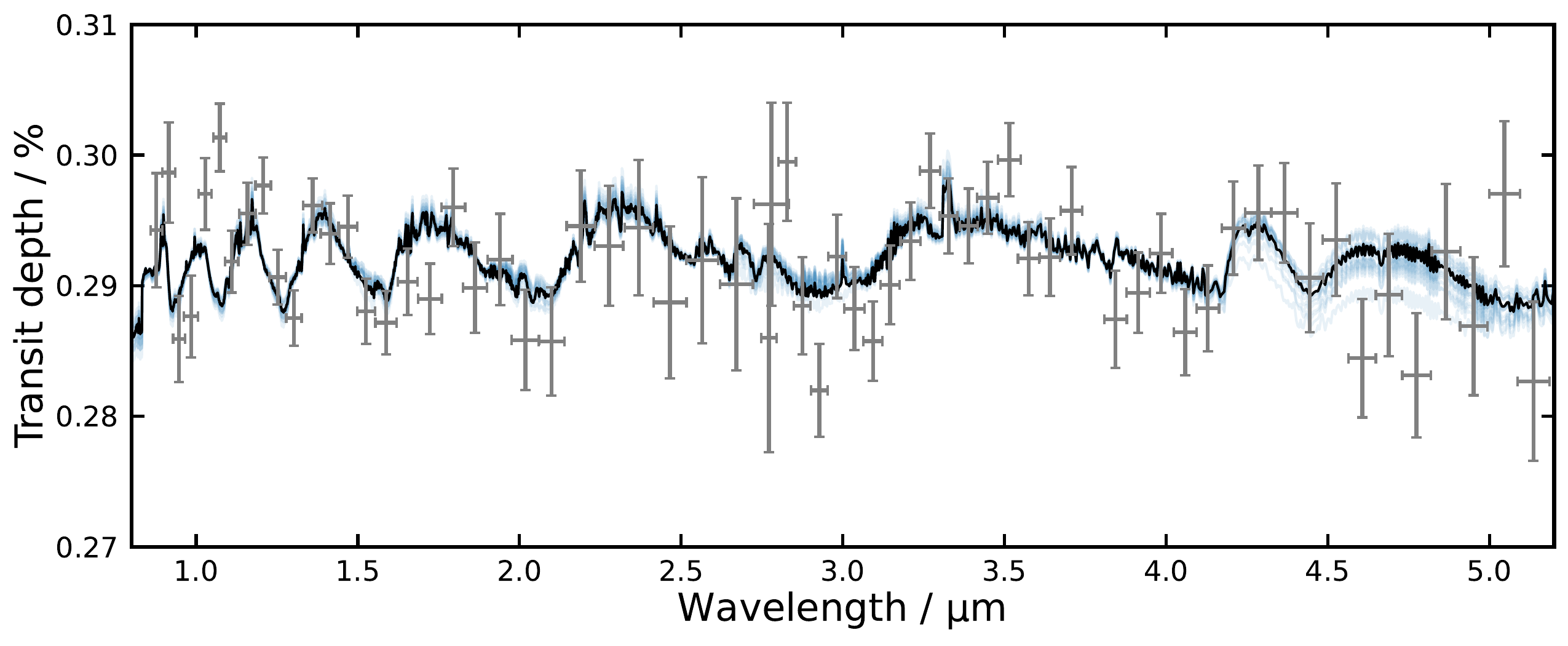}
    \caption{Best fitting model transmission spectra of the atmospheres in the magma ocean scenario. The transmission spectra are generated, using petitRADTRANS, from the output photochemical-kinetics models when taking the atmosphere-magma ocean system output as a lower boundary condition (step 4, figure \ref{fig:method_schematic}). Data points from the observed transmission spectrum of K2-18b \citep{Madhusudhan2023} are plotted with their associated error estimates. The best fitting model in the magma ocean scenario is shown in black and all models that agree with the observational data within 3$\sigma$ are shown in blue.}
    \label{fig:spectra}
\end{figure}

\begin{figure}[htbp]
    \centering
    \includegraphics[width=0.8\columnwidth]{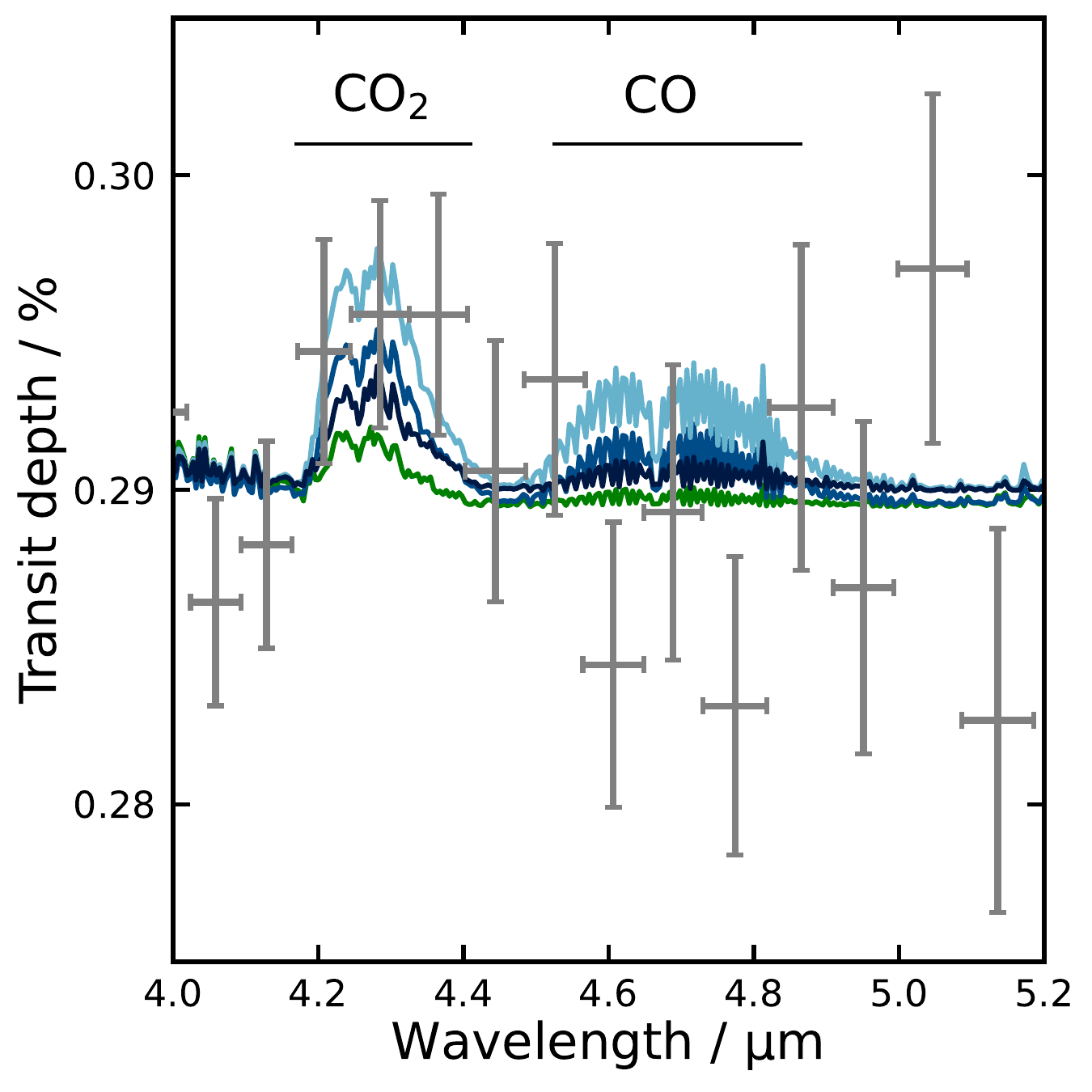}
    \caption{Transmission spectra within 4.0 - 5.2\,$\mu$m for four models demonstrating a range of \ce{CO2}:\ce{CO} ratios from the magma ocean scenario. Observational data from the observed transmission spectrum of K2-18b are shown with associated errors.}
    \label{fig:CO2_CO}
\end{figure}

A set of the resulting atmospheres in the magma ocean scenario are consistent with the full transmission spectrum of K2-18b observed by JWST \citep{Madhusudhan2023} within $\sim\,3\sigma$; this represents a suitable goodness of fit for a self-consistent physical model as compared to an optimised free-chemistry retrieval (Figure \ref{fig:spectra}). The transmission spectra shown in figure \ref{fig:spectra} are calculated with petitRADTRANS from the chemical profiles output by the photochemical-kinetics model (step 4 in figure \ref{fig:method_schematic}). We find that the atmospheric profiles output from the photochemical-kinetics model are relatively insensitive to the vertical transport parametrisation and photochemistry, however the condensation of water at $\sim$ 1\,bar and the thermochemical production of \ce{CO2} and \ce{CO} can significantly influence some of the resulting model spectra. Notably, water condensation decreases the mean molecular weight of the upper atmosphere and removes the presence of water absorption features from the spectrum.
\\\\
In all suitably well-fitting cases, the model spectra show evidence of \ce{CH4} features across the wavelength range $<4\,\mu$m, and evidence of \ce{CO2} at $\sim 4.3\,\mu$m (Figure \ref{fig:spectra}), in agreement with the previously reported free-chemistry retrieval analysis \citep{Madhusudhan2023}. The precise fit to the observed data would be further improved by tuning the temperature and mean molecular weight, and including additional fitting parameters such as cloud and haze scattering parameters and/or spectral features, cloud coverage fractions over the limb, etc., that are generally employed in free-retrieval analyses and allow increased degrees of freedom in improving goodness of fit. Optimising the spectral fit for the magma ocean scenario presented here is beyond the scope of this work; we instead suggest that this self-consistent magma ocean model can produce a qualitatively similar transmission spectrum to that observed for K2-18b, and those hypothesised for Hycean planets generally.
\\
\\
The presence of \ce{CO2} at $\sim 4.3\,\mu$m in the model spectra from the magma ocean scenario is generally also accompanied by a lower amplitude \ce{CO} feature at $\sim 4.7\,\mu$m, demonstrated with four representative cases in Figure \ref{fig:CO2_CO}. The transmission spectrum of K2-18b obtained with JWST shows significant evidence for the presence of \ce{CO2} at $\sim 4.3\,\mu$m and a preference for no \ce{CO} in the spectrum, with a reported mixing ratio of $\lesssim$ 10$^{-3}$ at the 95\,\% confidence level \citep{Madhusudhan2023}. The observational data longward of $4.3\,\mu$m have larger associated errors, making it the most weakly constrained region of the spectrum. Likewise, in the fit for the magma ocean scenario here, the constraining power in the wavelength region of the \ce{CO} feature is less than that of the \ce{CO2} feature, and thus the best-fitting physical models are those that show significant evidence of \ce{CO2} and a moderate accompanying \ce{CO} feature. Deeper observation of this spectral region in the future is desirable in order to confidently diagnose the presence or absence of observable \ce{CO}.

\section{Discussion \label{sec:discussion}}

\subsection{Tracers of solubility equilibria}

Our results demonstrate an important caveat to the use of \ce{NH3} in constraining the interior structure of Mini-Neptune exoplanets, such as K2-18b: if mass radius estimates permit internal structures that place \ce{H2} above a molten silicate surface, \ce{NH3} cannot be employed to break the degeneracies. This is because the solubility equilibria for ammonia of both the magma ocean scenario and the water-world scenario are qualitatively similar to each other, whilst being distinct from the thermochemical equilibrium resulting from a Neputune-like interior structure. Thus, alternative mutually exclusive chemical tracers of the presence of a water ocean versus a magma ocean should be sought so that future observations can distinguish these potential scenarios.
\\\\
One such possible tracer, and source of potential misfit of the magma ocean scenario with the observed spectrum of K2-18b, is the co-existence of \ce{CO2} and \ce{CO}. In the atmospheres that we have modelled here, the transmission spectra at wavelengths $>4\,\mu$m can range from being flat and featureless to showing \ce{CO2} and \ce{CO} features at varying amplitudes. The amplitude of the \ce{CO} feature always remains less prominent than that of \ce{CO2}, which is also predicted for the proposed class of Hycean planets generally \citep{Madhu-Hycean-2023}.  However, the presence of \emph{any} \ce{CO} is more uncertain in the observed spectrum of K2-18b specifically. While the observed spectral region $>4.5\,\mu$m does not suggest evidence of \ce{CO} absorption, the data quality are not currently sufficient to rule out the presence of \ce{CO}, and future observations could target this region, or other wavelength regions where \ce{CO} should be detectable. 
\\\\
It is possible however, that even the presence of \ce{CO2} and \ce{CO} are insufficient to distinguish between the water world and magma ocean scenarios. First, the stratospheric temperatures may be greater than those used in the present study \citep[from][]{Benneke2019}. Higher atmospheric temperatures would increase the amplitude of spectral features which would result in taller \ce{CO2} features than those modelled here, and, in combination with a cloud deck, potentially obscured \ce{CO} features. This could be confirmed in the future with self-consistent modelling studies of the coupled climate-chemistry system of K2-18b with an underlying magma ocean. Second, different regions of the physical and chemical parameter space for the atmosphere-magma ocean system, beyond the range that we have explored in this study, may result in differing relative abundances of atmospheric \ce{CO2} and \ce{CO} in the deep atmosphere in contact with the magma ocean \citep[e.g.,][]{Yu2021,Hu2021,fortney2020}. Finally, it is possible that there may be chemical pathways which increase the efficiency of the oxidation of \ce{CO} to \ce{CO2} in the atmosphere that are not currently considered in our photochemical-kinetic reaction network, such as those hypothesised in the atmosphere of Venus to explain the outstanding \ce{CO2}-stability and \ce{O2}-overabundance problems \citep{Marcq2018}.
\\\\
The atmospheric chemistry and solubility equilibria of \ce{CO2} and \ce{CO} in both the magma ocean and water world scenarios deserve further investigation, as they will be crucial context for any potential biosignature detections in the future. In the Hycean hypothesis, the \ce{CO2} flux at the surface-atmosphere boundary is considered to vary based on the planet's bulk carbon composition, the degree of carbon sequestration in the core, the pH of the proposed ocean water, wet deposition via the rainout of \ce{H2O}, and the potential presence of carbon-fixing life \citep{Hu2021,Madhu-Hycean-2023}. Most notably, if a Hycean exoplanet with ocean dwelling life drew down \ce{CO2} via its metabolism to the point of creating an atmospheric transmission spectrum that appears \ce{CO2}-depleted and thus flat between 4 - 5\,$\mu$m, then this will be indistinguishable from a magma ocean scenario (Figure \ref{fig:CO2_CO}), leading to potential false positive biosignature detections in the future.

\subsection{Ocean or magma ocean?}

The surge of interest in the hypothesis that stable liquid water oceans could exist on mini-Neptune exoplanets stems from the initial prediction that K2-18b receives an instellation flux that places it within the liquid water habitable zone. Initial estimates of the location of the inner edge of the liquid water habitable zone for \ce{H2}-dominated atmospheres, using 1D radiative-convective equilibrium models, required that K2-18b have a thin ($\lesssim 1$\,bar) atmosphere to prevent the onset of runaway greenhouse and a supercritical steam atmosphere \citep{Hu2021,Scheucher2020}. Since K2-18b sits at the low density end of the 1.7 -- 3.5\,R$_{\oplus}$ (mini-Neptune) population, such a scenario requires that K2-18b must have a particularly small core mass fraction.  Such an interior structure may require fine-tuning from a planet formation perspective \citep{rogers2011_apj,lee2016_apj}. More recently, the inner edge of the liquid water habitable zone under a $\sim 1$\,bar \ce{H2}-dominated atmosphere was revised to lie at 0.280\,AU from K2-18 \citep{Innes2023}, significantly beyond the measured orbital separation of K2-18b at 0.159\,AU \citep{Benneke2019}. In this new estimation, K2-18b could only maintain a stable liquid water ocean, not only with a small core mass fraction, but also requiring a cloud layer capable of raising the planetary albedo to $\gtrsim 0.7$ permanently, without simultaneously inducing additional warming from aerosol absorption or the scattering greenhouse effect \citep{theBible}.
\\\\
Such a scenario for K2-18b is not implausible: it has been demonstrated that rocky planets with water oceans may potentially maintain stable surface water inside of the inner edge of the habitable zone due to climate feedbacks of substellar clouds, provided that they are sufficiently slowly rotating planets \citep{2013ApJ...771L..45Y,2023NatAs...7.1070Y,Way2020}. The Hycean scenario, however, may not be possible under the hot starting conditions from which the planet would evolve \citep{2018A&A...610L...1V,2018ApJ...869..163V,2018ApJ...854...21C,2020PNAS..11718264K,2022NatAs...6.1296K}. In addition, cloud redistribution from dayside to nightside can reverse their cooling effect to net warming \citep{2021Natur.598..276T,Turbet2023}. The magma ocean scenario presented here may thus be more commensurate with planet formation and climate evolution models \citep{2023ASPC..534.1031K,2023ASPC..534..907L}. Confirming or refuting the magma ocean scenario for K2-18b will hold important implications on the potential for sub-Neptune and super-Earth exoplanets to maintain habitable conditions inside of the liquid water habitable zone \citep{2021ApJ...914...84A,2021ApJ...922L...4D,Innes2023,Pierrehumbert2023}, building an important foundation for demographic tests of the runaway greenhouse transition \citep{2023arXiv230904518S}.

\section{Conclusions \label{sec:conclusions}}

Interior structure and climate models permit mini-Neptunes to have \ce{H2}-rich envelopes overlying molten silicate surfaces.  Our models demonstrate that such planets will have detectable \ce{CO2} and low \ce{NH3} abundances in their atmospheres.  This results from the high solubility of nitrogen in molten silicate at reducing conditions; the conditions that likely prevail if a planet has an \ce{H2}-dominated envelope. Comparing this model to recent JWST spectra of K2-18b demonstrates that self-consistent physical models of the magma ocean scenario can explain the spectra to within $\sim$3$\sigma$; providing qualitatively as consistent an explanation for the data as the Hycean (i.e., liquid water ocean) scenario.  Distinguishing between the water and magma ocean scenarios on mini-Neptunes may be possible through deeper observations in the $>4$\,$\mu$m region, via quantification of the atmospheric \ce{CO2}/\ce{CO} ratio: magma ocean scenarios, in the conditions we have explored, generally have lower \ce{CO2}/\ce{CO} ratios than free chemistry retrievals infer from the present data.  Developing clear disambiguating atmospheric tracers for the presence of liquid water versus magma oceans is key in our quest of finding potentially habitable worlds amongst the exoplanet population.

\section*{Acknowledgements}
We thank an anonymous reviewer for helpful comments allowing us to clarify the work presented. We thank Edwin Kite for pointing out an inconsistency on an earlier version of the y axis of Figure 2.  S.J. thanks the Science and Technology Facilities Council (STFC) for the PhD studentship (grant reference ST/V50659X/1). D.J.B. carried out this work within the framework of the NCCR PlanetS supported by the Swiss National Science Foundation under grants 51NF40\_182901 and 51NF40\_205606. T.L. was supported by the Branco Weiss Foundation, the Alfred P. Sloan Foundation (AEThER project,  G202114194), and NASA’s Nexus for Exoplanet System Science research coordination network (Alien Earths project, 80NSSC21K0593).

\bibliography{refs}{}
\bibliographystyle{aasjournal}

\end{document}